# Teaching students about informatics and astronomy using real data for detection of asteroids

**A L Boldea[1,2] and O Vaduvescu [3,4,5]**

[1]National Institute for Nuclear Physics and Engineering, P.O.BoxMG-6, RO-077125 Bucharest-Magurele, Romania, E-mail: alinusha_b@yahoo.com

[2]University of Craiova, A.I.Cuza 13, RO-200385, Craiova, Romania

[3] Isaac Newton Group of Telescopes (ING), Apartado de Coreos 321, E-38700 Santa Cruz de la Palma, Canary, Islands, Spain, E-mail:ovidiu.vaduvescu@gmail.com

[4] Institut de Mécanique Céleste et de Calcul des Éphémérides (IMCCE) CNRS-UMR8028, Observatoire de Paris, F-75014 Paris Cedex, France

[5] Instituto de Astrofisica de Canarias (IAC), vía Láctea s/n, E-38200 La Laguna, Tenerife, Spain

**Abstract**. In this paper we approach the astronomy teaching process for the students in computer sciences through the controlled investigation method on real astronomical data, using data reduction and quality control of the astrometry of near-Earth asteroids. The method used the data collected on the Isaac Newton Telescope (INT) located at the ORM observatory on the island of La Palma in the Spanish Canary Islands and was successfully tested on a group of students in the second-year of study.

## 1. Introduction

In order to construct an accurate picture of astrophysics and to gain an insight into the astronomic phenomena, new models were required. The development of these models was based, from chronological standpoint, on the discovery of new data. The enhancement of knowledge creates new models and paradigms. This was caused by the huge technical and technological evolution (particularly the information technology which made quick strides in the field of calculation of high performance thus achieving progress in the elaboration of algorithms, theories and models ([1], [2], [3]).

Three technological fields of the astronomy (namely telescopes, detectors and computers) constantly recorded, over time, significant advances. Consequently, an increasing amount of data has been generated. For this massive volume of data to be properly handled it will be necessary to augment the skilled force in the areas of computational and experimental astronomy ([4], [5]).

Nowadays, the science teachers guide their students through the path of modern astronomical concepts, of evidence-based explanations and of acquiring the competencies and the knowledge needed to apply the complex technologies in the field.

In the past, observational astronomy was based on watching the sky with the naked eye. People learned about the stars and the planets just by looking at the sky. Thus, were revealed the first five of our planets (Mercury, Venus, Mars, Jupiter, and Saturn). Today, the field of observational astronomy saw a rapid digitization which helps to accurately analyze the data obtained.

The students in astronomy should develop not only an understating of the "big picture" but also the knowledge that allows them to master the variety of technological instruments. However, in order to help them to acquire these skills and capabilities, non-classical method of teaching (brain-storming, problem-solving method etc.) seem the most appropriate to bring students directly into contact with real-world data, and to integrate data analysis.

This paper aims at presenting the results of a scientific-methodical training experiment involving the students in computational astronomy, related to processing real packages of astronomical images obtained by the Isaac Newton Telescope (INT) from La Palma (Canary Islands) within the international frame of the EURONEAR (the European Near Earth Asteroids Research) project ([6]). The EURONEAR consortium includes *inter alia* 3 Romanian institutions (AIRA Bucharest, University of Craiova and Technical University Cluj-Napoca).

The main methods used in this teaching-learning process were the investigation (used to learn the background noise reduction methods) and the discovery (used to identify and classify moving celestial bodies, primarily asteroids or possible comets).

## 2. The metodical experiment: the use of the investigation and discovery methods to teach computational astronomy

The *investigation learning method*, considered as a form of guided inquiry method, was described in cognitive learning sciences as the process of finding solutions to a problem by exploring alternative options, using methods that have already been learned by the students, under the guidance of a teacher [7]. The teaching strategies generally include a number of steps:
- cognitive preparation (verification of previous skill and knowledge),
- formulation of the problem,
- exploration (search for a solution, experimentation),
- analysis of first results and explanations,
- group communication and evaluation of the information,
- evaluation by the teacher
- applications to similar cases ([8]).

On the other part, the *discovery learning method,* as it was formalized by Bruner [9], is a constructivist learning theory applying in problem solving situations where the students use their knowledge and the accessible sources of information in order to discover facts, relationships or new knowledge by their own. This paradigm of learning generates more models of education that include: guided discovery, simulation based learning, incidental learning and others [10].

Our goal was to combine the investigation method (applied in the learning process of some key concepts in graphic image processing) and the information obtained through the discovery method (applied in order to learn the basic concepts of the asteroid astrometry).

The methodical research hypothesis in the context of our experiment was: *The application of investigation and the discovery method in teaching and learning of computational astronomy in general and astrometry, in this particular experimental case, for a group of students in the second-year of study, by using some specialized softwares and real images packages obtained with a powerful telescope, can provide: the increase in the volume of the assimilated information; the reduction in the training time; the development of the competencies needed for using specialized IT technologies in the study of astronomy and, more important, the* formation *of the* research competency *in the field of astronomy.*

The following objectives have been pursued:
- determination of the effectiveness of the investigation method with a view to familiarizing the students with specialized tools for processing graphic images, particularly comparative analysis of the methods for reducing the background noise from astronomical photographs as well as the formation of competencies in analyzing and processing the data in computational astronomy,
- the verification of the usefullness of the discovery learnind method by using real data, with a view to training the future researchers in astronomy through formation of competencies that will enable them to track and to identify near-Earth asteroids.

The methodical hypothesis verification, according the set objectives, used a teaching research approach based on the comparative study of the evolution of the training process on small groups of students, one of which benefited only of a traditional training in Astrometry (control sample).

In order to achieve the objectives of the investigation, the following research methods have been employed: pedagogical observation; pedagogical experiment; the test of the level of knowledge assimilation of the experimental group in relation to experimental and control lessons; the test of the competencies acquired by the students by using real data packages and statistical data analysis.

The study was performed in the late spring of 2016 on a group of students in the second year of study in Computer Sciences at the University of Craiova, also enrolled in the experimental course of Computational Astronomy.

The experimental course of computational astronomy included several modules dedicated to the use of various applications software dedicated to processing the images taken by large telescopes and the identification of the various bodies that appear in them. For the experimental module described in this paper, we used the specialized software program called *Astrometrica* ([11]), presented in the Section 4.

The students were divided into three working groups, each group having prior some knowledge of Astronomy acquired in the previous modules (*e.g.* relative and absolute celestial coordinates, positioning of the main astronomical bodies in the Solar System, relative and absolute magnitude). In addition, at the beginning of the experimental approach, the students was supposed to have been trained to use Java programming language and to posses some basic elements of computer graphics.

For performing the experiment it was used the *2002RP28a* data package of pictures, obtained at Issac Newton Telescope (INT), with the Astronomical Observatory code 950, in La Palma (the Spanish Canary Islands) on March 30, 2016. The aforementioned package consisted of groups of six images each, taken with the four CCD cameras of the telescope, at a time interval of a few minutes between two consecutive images. The images were previously reduced and converted into FITS format.

In the first stage, the students in the two study groups, except the control group, were asked to use a local software utility that allows the combination of multiple methods of graphics processing to reduce the background noise from the studied images. This step is considered necessary because the brightness of the asteroids is generally very low and thus, it is often difficult to identify them.

The most popular techniques for reducing the background noise from the photos, obviously, are not directly applicable on astronomical images, as they pose the risk of eliminating from image the objects which have very low brightness; therefore the students were set the objective to determine a combination of the filters that have been used in order to optimize the image quality without definitely blurring the possible traces of asteroids. At this step, the students were evaluated by their ability to produce final images with comparable low level of background noise, by reproducing their best combination strategy. The control group do not participate at this stage of the experiment.

In the second stage, all three groups of students had to identify, by using the Blinking option from *Astrometrica* software, as many traces of asteroids as possible from the analyzed photo package.

### 3. The reduction of the images noise. Investigation approach

A first step of this experiment involved the use of a personal software package implemented in Java at University of Craiova – the *Java digital image processing* (*JDIP*) shortly presented in the Appendix - in order to eliminate the background noise from each of the images as much as possible, without eliminating all the small moving objects. The package of pictures composed by six images taken in a short time (typically 15 minutes) with the Isaac Newton Telescope (2.5 m diameter), pointing to the same direction during few consecutive short exposures (typically one minute).

A graphic noise appears into a CCD image due to various atmospheric bright phenomena, such as the reflection of light from the environment, the lightning or convection phenomena, and cosmic radiation. These disturbances superimposed over the picture, altering the image in an undesirable way. Therefore, these disturbances largely depend on the transmission and reception procedures. Consequently, we cannot speak of a single and universal algorithm to eliminate these disturbances, because these algorithms need to be adapted to both the types of the noises and the characteristics of the non-altered objects from the image.

The graphical noise reduction methods used classically the linear filter methods based on a convolution with a kernel on the form of the Sharpen kernel, the average and median filter (the Blur kernels), average weighted Gaussian filter or wedge recognition. None of these methods was successfully performed directly on images taken by a CCD camera of a telescope.

The investigation method approach to the problem of reducing the background noise from digital CCD camera's images followed the main steps outlined in the previous section. After a period of familiarization with the software tool JDIP and a brief review of methods of images processing through the masks of convolution, students were asked to identify and test the effect of variants of convolution different types of convolution kernel, combined with other visual effects implemented (Posterization, Threshold, Color Balance), in order to obtain the best noise reduction.

After obtaining a positive result in filtering the image, each student was asked to analyze his results in write and to compare them with the results of the other students from the same group. This step of analyse and confrontation had the goal to refine the best combination of methods chosen by each student.

Finally, each student has applied its own strategy for reducing noise from images on the entire package of pictures. The evaluation of the teacher included checking the ability of students to reproduce with the equivalent level of quality the results obtained with the rest of the CCD's photos.

Of the two participating groups of students, 80% were able to successfully replicate an effective strategy for noise reduction for the other photos from the analyzed package, while 20% suggested strategies that proved effective only for 1-2 images.

By comparing the final methods proposed by the two groups of students, it can be concluded that: the best filtering result was obtained by a combination of three basic operations. First, the original image is submitted to a reduction of the weigh (bright) values in the HSB representation with a level *w* between 25 % and 29 %. A high level reduction produces a loss of the contrast of the image; a lowest level of *w* produces no visible effect.

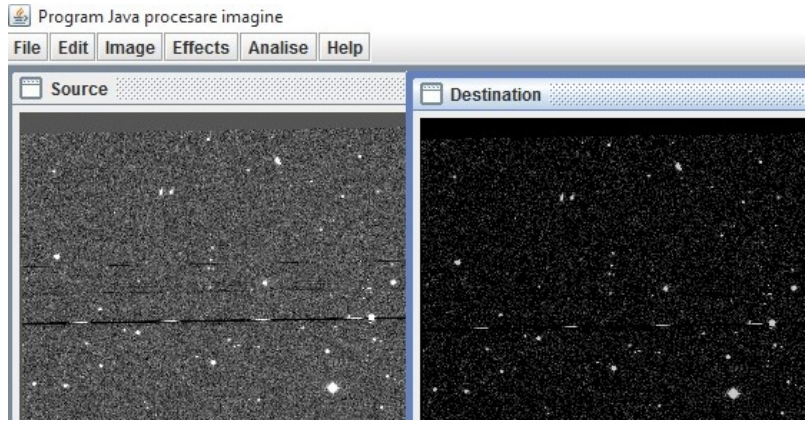

Figure 1. The effect of the reduction of the weigh value in the HSB representation over the intensity of the gray levels

The second step is a posterization of the image using 8 levels of grayscale. Gray value for each pixel has been converted to the nearest level of eight equidistant levels.

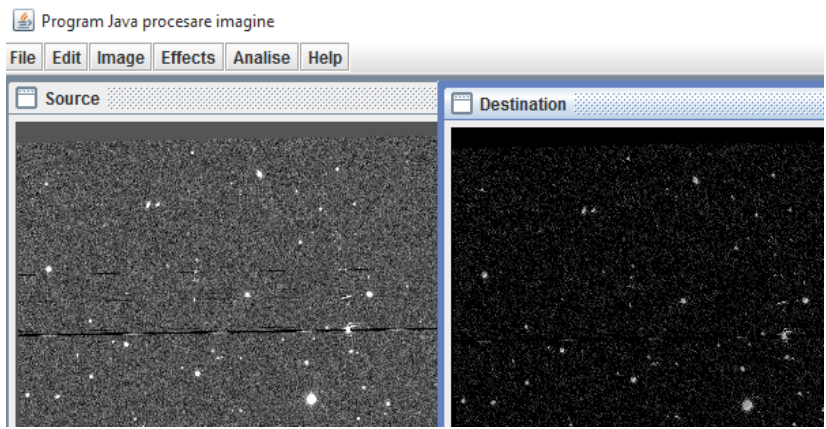

Figure 2. The Posterization effect. A small reduction of noise is visible

The final noise reduction used a filter mask of the form:

$$h = \frac{1}{2}\begin{bmatrix} -a & 0 & -b \\ 0 & 2 & 0 \\ b & 0 & a \end{bmatrix}$$

(5)

where *a* and *b* (*a, b > 0*) are control parameters. The best effects were obtained for the values *a = 1, b = 0.4*

The convolution was necessary to amplify the luminosity of the astronomical objects from the image (stars, asteroids, comets) and to dilute background variation.

*Teaching students about informatics and astronomy using real data for detection of asteroids*

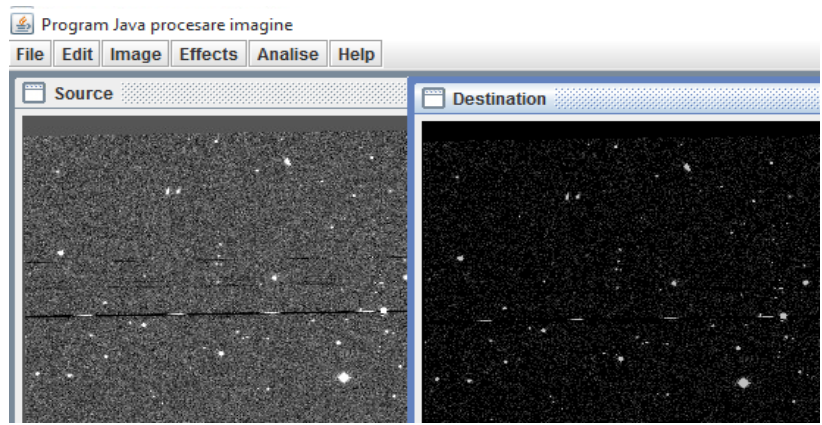

Figure 3. The final effect

## 4. The discovery approach to the identification of asteroids

After processing the entire package of images, the students of all the three groups were asked to use the specialized software *Astrometrica* in order to identify the moving objects. The first two groups have used their own processed images after the noise reduction; they worked individually and they only addressed the teacher for a feedback. The control group used the initial image before the noise reduction and worked under the direct supervision of the teacher. The number of moving objects identified and the average time for obtaining the astrometric data (the universal coordinates) for a supposed asteroid, through the facilities of provided by the software tool, constituted the control parameters for evaluating the result of the methodical experiment.

*Astrometrica* is a tool application that permits automatic images calibration (by superposition of fixed objects identified as stars), automatic identification of reference stars (using a direct connection to a database holding stellar catalogs) and also stacking multiple images to detect very faint asteroids. *Astrometrica* allows the user to measure the position of moving sources and create a report file to be sent to Minor Planet Centre (MPC) - the main database of asteroids in Solar System [12].

The search of an asteroid passes through the "false animation" of the package of pictures by using a technique called "blinking", chronological display of the individual images in succession, recalibrated until the known stars are superposed. The obtained animation effect, called *blinking,* allows the visual identification of moving objects from the images, moving objects that are not stars or planets. For each of these observed objects, the students extracted its successive coordinates, using an Astrometrica facility, and generate a report in MPC format.

The method is laborious and requires a careful study of each fragment of the picture, the verification of the observations by the teacher of students and the validation of these observations. We were interested in comparing the observation capacities and the correct identification of asteroids for each group of students.

An additive summary of the students' observations is showed in the Table 1.

**Table 1**  The centralized identification data for the supposed new asteroids, using the MPC reports

| Temporary name | Date (julian format) | Right ascension | Declination | Apparent Magnitude |
|---|---|---|---|---|
| EUVB001 | 2016 03 30.01359 | 11 49 25.50 | +07 18 58.6 | 20.1 |
| EUVB001 | 2016 03 30.01497 | 11 49 25.43 | +07 18 59.2 | 20 |
| EUVB001 | 2016 03 30.01634 | 11 49 25.36 | +07 18 59.9 | 20.1 |
| EUVB001 | 2016 03 30.01772 | 11 49 25.29 | +07 19 00.6 | 20 |
| EUVB001 | 2016 03 30.01910 | 11 49 25.22 | +07 19 01.2 | 19.9 |
| EUVB001 | 2016 03 30.02046 | 11 49 25.15 | +07 19 01.8 | 20 |
| Temporary | Date | Right | Declination | Apparent |

*Teaching students about informatics and astronomy using real data for detection of asteroids*

| name | (julian format) | ascension | | Magnitude |
|---|---|---|---|---|
| EUVB002 | 2016 03 30.01497 | 11 49 08.22 | +07 11 15.3 | 21.5 |
| EUVB002 | 2016 03 30.01634 | 11 49 08.17 | +07 11 16.0 | 21.2 |
| EUVB002 | 2016 03 30.01772 | 11 49 08.10 | +07 11 16.4 | 21.3 |
| EUVB002 | 2016 03 30.01910 | 11 49 08.05 | +07 11 16.9 | 21.6 |
| EUVB002 | 2016 03 30.02046 | 11 49 07.99 | +07 11 17.5 | 21.3 |
| EUVB003 | 2016 03 30.01359 | 11 48 53.80 | +07 15 28.4 | 21.7 |
| EUVB003 | 2016 03 30.01497 | 11 48 53.75 | +07 15 28.1 | 21.9 |
| EUVB003 | 2016 03 30.01634 | 11 48 53.63 | +07 15 28.2 | 21.3 |
| EUVB003 | 2016 03 30.01772 | 11 48 53.53 | +07 15 28.1 | 21.7 |
| EUVB003 | 2016 03 30.01910 | 11 48 53.48 | +07 15 28.2 | 21.5 |
| EUVB003 | 2016 03 30.02046 | 11 48 53.35 | +07 15 28.0 | 21.7 |
| EUVB004 | 2016 03 30.01359 | 11 48 52.78 | +07 09 21.7 | 20.7 |
| EUVB004 | 2016 03 30.01497 | 11 48 52.77 | +07 09 22.3 | 20.7 |
| EUVB004 | 2016 03 30.01634 | 11 48 52.73 | +07 09 22.9 | 20.8 |
| EUVB004 | 2016 03 30.01772 | 11 48 52.69 | +07 09 23.6 | 20.9 |
| EUVB004 | 2016 03 30.01910 | 11 48 52.67 | +07 09 24.4 | 20.7 |
| EUVB004 | 2016 03 30.02046 | 11 48 52.63 | +07 09 25.3 | 20.8 |
| EUVB005 | 2016 03 30.01359 | 11 48 37.81 | +07 18 52.8 | 17.3 |
| EUVB005 | 2016 03 30.01497 | 11 48 37.79 | +07 18 53.4 | 17.4 |
| EUVB005 | 2016 03 30.01634 | 11 48 37.76 | +07 18 54.1 | 17 |
| EUVB005 | 2016 03 30.01772 | 11 48 37.72 | +07 18 54.9 | 16.9 |
| EUVB005 | 2016 03 30.01910 | 11 48 37.70 | +07 18 55.6 | 17.1 |
| EUVB005 | 2016 03 30.02046 | 11 48 37.65 | +07 18 56.5 | 16.9 |
| EUVB006 | 2016 03 30.01359 | 11 48 39.90 | +07 14 48.2 | 21.4 |
| EUVB006 | 2016 03 30.01497 | 11 48 39.83 | +07 14 48.5 | 21.3 |
| EUVB006 | 2016 03 30.01634 | 11 48 39.77 | +07 14 49.0 | 21.2 |
| EUVB006 | 2016 03 30.01772 | 11 48 39.71 | +07 14 49.1 | 21.1 |
| EUVB006 | 2016 03 30.01910 | 11 48 39.65 | +07 14 49.7 | 21.6 |
| EUVB006 | 2016 03 30.02046 | 11 48 39.59 | +07 14 49.9 | 21.4 |
| EUVB007 | 2016 03 30.01359 | 11 48 34.47 | +07 14 30.5 | 20 |
| EUVB007 | 2016 03 30.01497 | 11 48 34.39 | +07 14 30.8 | 20 |
| EUVB007 | 2016 03 30.01634 | 11 48 34.31 | +07 14 31.1 | 20 |
| Temporary | Date | Right | Declination | Apparent |

| name | (julian format) | ascension | | Magnitude |
|---|---|---|---|---|
| EUVB007 | 2016 03 30.01772 | 11 48 34.23 | +07 14 31.3 | 20 |
| EUVB007 | 2016 03 30.01910 | 11 48 34.16 | +07 14 31.6 | 19.9 |
| EUVB007 | 2016 03 30.02046 | 11 48 34.07 | +07 14 31.9 | 19.9 |
| EUVB008 | 2016 03 30.01359 | 11 48 37.34 | +07 09 52.7 | 18.7 |
| EUVB008 | 2016 03 30.01497 | 11 48 37.32 | +07 09 53.3 | 18.7 |
| EUVB008 | 2016 03 30.01634 | 11 48 37.29 | +07 09 54.0 | 18.6 |
| EUVB008 | 2016 03 30.01772 | 11 48 37.24 | +07 09 54.8 | 18.7 |
| EUVB008 | 2016 03 30.01910 | 11 48 37.23 | +07 09 55.6 | 18.6 |
| EUVB008 | 2016 03 30.02046 | 11 48 37.18 | +07 09 56.4 | 18.6 |
| EUVB009 | 2016 03 30.01359 | 11 48 25.64 | +07 09 05.4 | 20.9 |
| EUVB009 | 2016 03 30.01497 | 11 48 25.64 | +07 09 05.3 | 20.8 |
| EUVB009 | 2016 03 30.01634 | 11 48 25.64 | +07 09 05.2 | 21.1 |
| EUVB009 | 2016 03 30.01772 | 11 48 25.63 | +07 09 05.4 | 20.8 |
| EUVB009 | 2016 03 30.01910 | 11 48 25.64 | +07 09 05.2 | 20.7 |
| EUVB009 | 2016 03 30.02046 | 11 48 25.65 | +07 09 05.2 | 20.8 |
| EUVB010 | 2016 03 30.02046 | 11 48 18.21 | +07 10 19.7 | 21.5 |
| EUVB010 | 2016 03 30.01359 | 11 48 18.22 | +07 10 19.7 | 21.4 |
| EUVB010 | 2016 03 30.01497 | 11 48 18.20 | +07 10 19.6 | 21.4 |
| EUVB010 | 2016 03 30.01634 | 11 48 18.19 | +07 10 19.8 | 21.4 |
| EUVB010 | 2016 03 30.01772 | 11 48 18.22 | +07 10 19.4 | 21.6 |
| EUVB010 | 2016 03 30.01910 | 11 48 18.16 | +07 10 19.6 | 21.8 |

On the original images, the students from the control group were able to visually identify only the objects EUVB001, EUVB004, EUVB005 and EUVB008 (see the figure 4 for the EUVB001 detection and computation of his astrometric coordinates with *Astrometrica* software). The use of the images with the background noise reduced by the innovative methods of the students from the last two groups proved to be clearly more useful for identifying the asteroids.

*Teaching students about informatics and astronomy using real data for detection of asteroids*

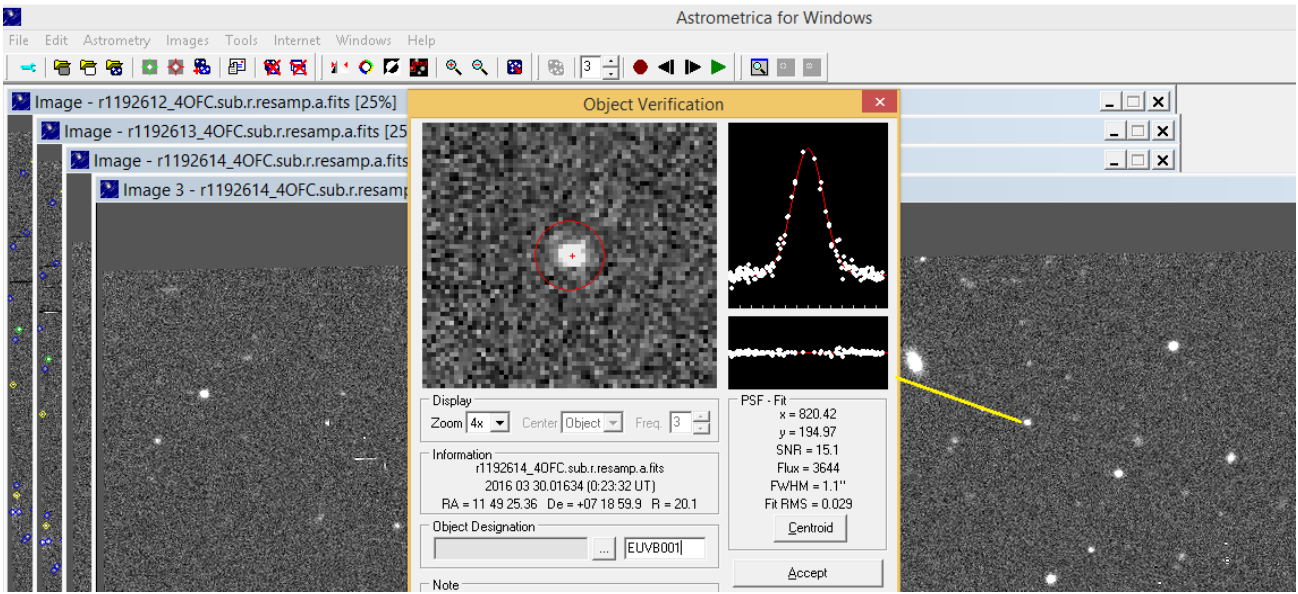

Figure 4. The EUVB001 asteroid detected with the Astrometrica software

After generating the MPC report and the photometry report for the moving objects, the students must performed a validation of the report through the Astro Check[1] tool of EURONEAR project [6]. The software tool returns a list of objects for which the computation of successive positions presents too big errors. In the case of the observations in Table 1, the results signaled possible errors of the correct value of declination for the second observation of EUVB003, the third observation of the EUVB009; the EUVB010 data are also rejected.

Figure 5. The validation of the MPC report. Last object is rejected for observation errors reasons

---

[1] http://www.euronear.org/tools/astchk.php

The next step was the identification of each asteroid observed in the MPC database. In order to do this, we used the MPChecker facility from the Minor Planet Center[2]. We present below a fragment of the MPChecker report for the EUVB001 object. Notice that the Checker computes the position for more that 930,000 know asteroids at the moment of the observation and extracts from the list, the asteroids that are visible in a limited range (less that few arcmin). Also, note that for a difference more that *10''* arcsec between the Right Ascension (*R.A.)* and the Declination (*Decl.)* of the observed position and the computed position of an asteroid from the MPC list, the observed object is considered as different one.

```
MPChecker/CMTChecker/NEOChecker/NEOCMTChecker

Here are the results of your search(es) in the requested field(s) (positions are determined from elements integrated to a nearby epoch) :

The following objects, brighter than V = 23.0, were found in the 7.0-arcminute region around the following observation: No known minor planets, brighter than V = 23.0

     EUVB001  C2016 03 30.01359 11 49 25.50 +07 18 58.6          20.1 R      950

Object designation         R.A.      Decl.     V     Offsets    Motion/min  Orbit  Further observations?
                          h  m  s    °  '  "         R.A. Decl. R.A.  Decl.        Comment (Elong/Decl/V at date 1)

         2010 SK35        11 49 19.9 +07 24 10  21.9  1.4W  5.2N  0.6-  0.2+   3o  Desirable between 2017 May 7-June 6. (
(427163) 2014 UQ189       11 49 46.4 +07 15 52  21.6  5.2E  3.1S  0.6-  0.3+   6o  None needed at this time.
         2002 RP28        11 49 22.5 +07 12 38  22.9  0.7W  6.3S  0.8-  0.5+   2o  NEO : Very desirable between 2017 May 7

Number of objects checked = 931637
```

Figure 6. The EUVB001 MPChecker report. The asteroids in the range of 7' from the observed position the 2016 mars 30.01359 Julian day

After the validation of the students' observations and the verification in the MPC database of asteroids, comets and minor planet, the result indicates that EUVB001, EUVB002, EUVB004, EUVB005, EUVB006 and EUVB008 may be considered as newly discovered asteroids; but series of new observations will be necessary in order to compute with sufficient accuracy the trajectories of these objects and to insert them into the databases of known asteroids. Only the EUVB007 object was identified as a known asteroid: the (308710) 2006 GO16 asteroid, discovered the 2006.04.02 by the Kitt Peak Observatory (see the figure 7). The EUVB003, EUVB009 and EUVB0010 presents some variation of the trajectory, possibly due to some errors of observation or to the influence of the background noise. The observation data for these three "objects" are considered to poor to be recoded as valid observation.

```
MPChecker/CMTChecker/NEOChecker/NEOCM

Here are the results of your search(es) in the requested field(s) (positions are determined from elements integrated

The following objects, brighter than V = 23.0, were found in the 7.0-arcminute region around the following observ

     EUVB007  C2016 03 30.01359 11 48 34.47 +07 14 30.5          20.0 R      950

Object designation         R.A.      Decl.     V     Offsets    Motion/min  Orbit
                          h  m  s    °  '  "         R.A. Decl. R.A.  Decl.

(308710) 2006 GO16        11 48 34.5 +07 14 30  20.6  0.0E  0.0S  0.6-  0.1+   8o
```

Figure 7. The EUVB007 MPChecker report

---

[2] http://cgi.minorplanetcenter.net/cgi-bin/checkmp.cgi

## 5. Conclusions

In order to measure the effectiveness of our experimental approach, we proceed to the comparison of the investigation method used for the two groups of students while applying the traditional teaching method for the third one. A short summary of the indicators is given below:

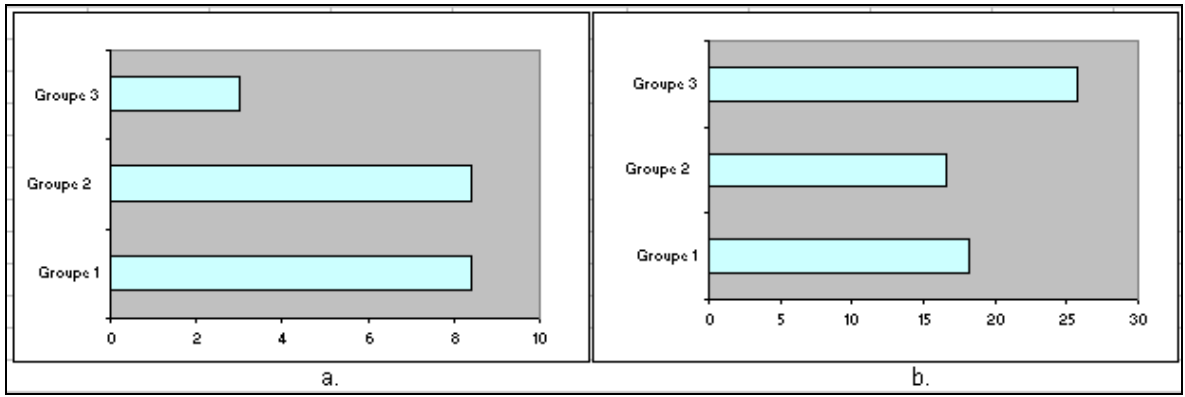

Figure 8. a. The average number of moving objects (supposed asteroids) discovered by all the groups.
b. The average time (in minutes) to discover a moving object

The ANOVA analysis shows, for a confidence level *alpha*=0.02, that the average number of identified objects and the average discovery time, present, from statistical point of view, significant differences between the first two groups and the third ($F_{values}$ were 40.50, and 11.804 respectively, for the two indicators, where $F_{critic}$=5.516).

Then we can appreciate that the methods of guided investigation and discovery were successfully used to initiate a direct approach of the students to one of the hottest topics in the research area of astronomy: the detection of new asteroids. The comparison between this method and traditional method showed a significant advantage of the first one. This approach enables the future researchers, while they are still at beginner's level, to discover and identify in short time some new astronomical objects.

In conclusion, the direct approach of the research area in astronomy using the investigation and discovery methods yielded good result in a very short period of time.

## Appendix. The description of the software application JDIP

*Java digital image processing* (JDIP) is image processing tool produced at the University of Craiova that we used in the methical experiment with the students. This software package allows the partial elimination of the galactic nebulae and galaxies, the environmental noise, as well as the extraction of the edges and "shapes" of the astronomical objects from the images (stars, asteroids, comets). JDIP has a standard interface similar to others graphical tools. Its main window contains a menu with the sections Files, Edit, Image, Effects, Statistics and About, each of them being divided into sub-menus, plus the graphical area, spitted in two parts: for the original image and the modified one. The tool was implemented in Java, with the package Matis. The first three menu sections permit to load or save a file, to undo the user modification of the images and to modify the orientation and the dimension of the image. The interesting part of the application is the section Effects.

JDIP use convolution mask to process the images, method that allow to partially filter the background graphic noise. In order to obtain meaningful effects, the most suitable convolution mask must be chosen, through successive attempts, by using variations of the parameters of convolution mask for the images studied.

The main images processing sections, implemented in this software, use, in essence, a convolution at local level with standard or user-defined kernel masks ([13]). The convolution formula expresses the modification of the value of each basic colors of each pixel from an image using a predefined kernel matrix $K = (k_{i,j})_{i=-q..q;\ j=-q..q}$, where *q* is the dimension of the neighborhood radius of a given pixel:

$$p^*_{a,b} = \frac{1}{f} \sum_{i=-q}^{q} \sum_{j=-q}^{q} k_{a+i,b+j} \, p_{a+i,b+j} \quad (1)$$

Here $p^*_{a,b}$ is the new value of one of the colors, associated with the pixel position *(a,b)*, depending on the old values of his neighborhood and *f* is the sum of all the elements of the kernel matrix (*f = 1* if this sum is null).

The Effects menu carries some classical graphic transformation, not directly useful for the processing of a CCD image. The first operation is image enhancement (Sharpen). The convolution kernel has the form:

$$K_1 = \begin{bmatrix} 0 & -s & 0 \\ -s & 1+4s & -s \\ 0 & -s & 0 \end{bmatrix} \quad (2)$$

Unfortunately, the Sharpen effect, ordinary understood as a contrary of the Blur effect, do not eliminate the noises from a CCD image, but amplifies them (figure A1).

The Edge Recognition was implemented using the convolution kernel ([13]):

$$K_2 = \begin{bmatrix} 0 & -1 & 0 \\ -1 & 4 & -1 \\ 0 & -1 & 0 \end{bmatrix} \quad (3)$$

The effect on the CCD image taken with a telescope was similar to the one obtained in the previous case.

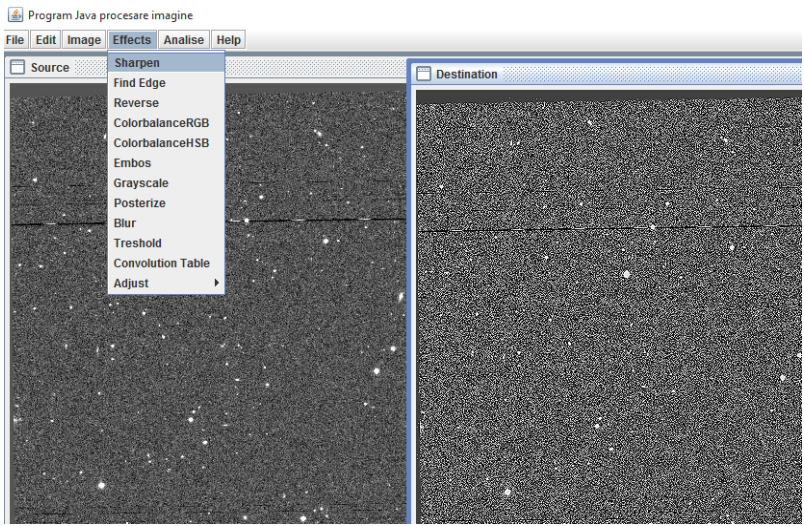

Figure A1. The effect of Sharpen convolution for a photo taken by a telescope

The application also allows the study of the Emboss effect, obtained with a convolution kernel of the form:

$$K_3 = \begin{bmatrix} -n & 0 & 0 \\ 0 & 1 & 0 \\ 0 & 0 & n \end{bmatrix} \quad (4)$$

where *n* is the magnitude of the effect.

The figure A2 presents the result of the direct application of the Emboss effect on a CCD image.

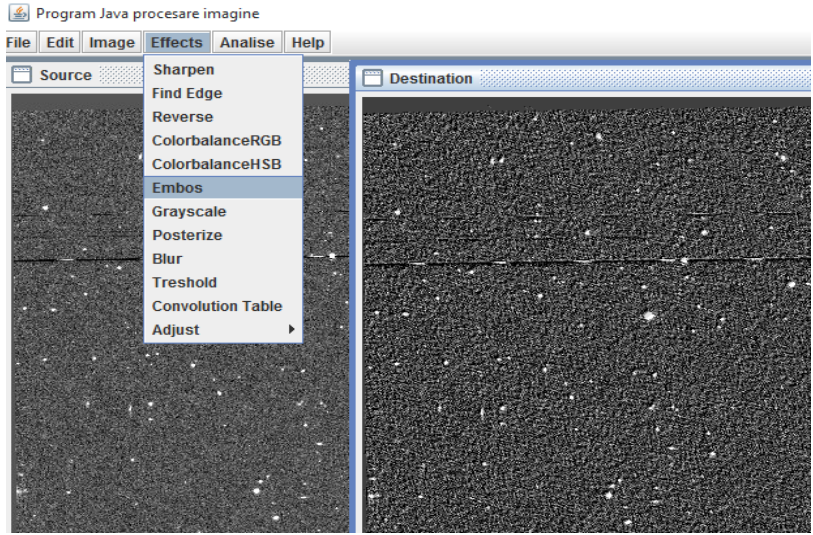

Figure A2. The effect of Emboss convolution for a CCD picture. The background noises becomes 3D

The application implements also the Blur effect (smoothening effect), the Posterization effect and the Black and White reverse effect. In order to finalize, the user has the possibility to define a 3x3 convolution kernel and to test the effect on the image, as presented in the figure A3.

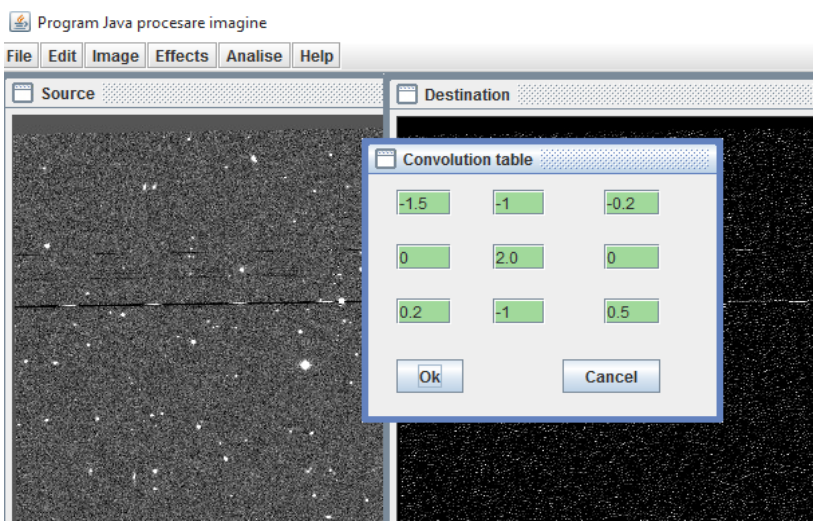

Figure A3. The convolution kernel defined by the user

We also implemented a method allowing for the adjustment of both the saturation and the weight value in the HSB color model of a digital image. The HSB color model is based on three components: Hue, Saturation and Brightness (weight). The application converts the image from RGB format (additive color model - Red, Green and Blue) to HSB format, operates the user modifications of the three aforementioned values and reconverts the image in the original form.